\documentstyle[12pt,aaspp4]{article}
\begin{document}

\title{Emission line imaging of QSOs with high resolution}

\author{J.B. Hutchings\altaffilmark{1}}
\author{S.L.Morris\altaffilmark{1}}
\author{D.Crampton\altaffilmark{1}}
 \affil{Herzberg Insitute of Astrophysics,
National Research Council of Canada,\\ Victoria, B.C. V9E 2E7, Canada}

\altaffiltext{1}{Observer with the Canada France Hawaii Telescope which
is funded by NRC of Canada, CNRS of France, and the University of 
Hawaii}

\begin{abstract}

  We report the first detection of emission line gas within the host  
galaxies
of high redshift QSOs. This was done using narrow-band imaging at the
redshifted wavelengths of [O~III] and H$\alpha$, using the PUEO  
adaptive optics
camera of the CFHT. The QSOs are all radio-quiet or very compact radio  
sources.
In all five observed QSOs, which have redshifts 0.9 to 2.4, we find  
extended
line emission that lies within 0.5" (a few Kpc) of the nucleus. The  
emission
(redshifted) equivalent widths range from 35 to 300A.  Where there is  
radio
structure, the line emission is aligned with it. We also report on  
continuum
fluxes and possible companions. Two of the QSOs are very red, and have  
high
resolved continuum flux. 

\end{abstract}


\section{Introduction and observations}

   We have previously reported broad-band (JHK) imaging observations of  
QSOs of
redshift 0.9 to 4.2 at resolutions $\sim$0.2 arcsec (Hutchings et al  
1999:
paper 1), which indicated extended or tidal structures in most of them.  

We also reported finding that the radio-loud QSOs appear to live in  
dense
environments of small companion galaxies whose colours suggest young
stellar populations. These earlier data, like those reported here, were  
taken 
with the PUEO adaptive optics system of the Canada France Hawaii  
Telescope.

   This paper describes observations of QSOs made with the same system 
using NIR narrow band filters to isolate the emission from the lines
of [O~III], and H$\alpha$ at the redshift of the QSOs. The idea
is to study the morphology of emission line gas in and around the QSOs,
and possibly to detect such emission from associated galaxies. Thus,  
this
investigation studies the gas in the environment of high redshift QSOs,
rather than the stellar content seen in the continuum imaging  
previously
reported. It is known in low redshift QSOs that extended emission line
gas is seen in and even beyond the stellar population, and is evidence
for stripping of gas in galaxy collisions, and illumination by the  
active
nucleus (Boroson, Persson, and Oke 1985,  
Stockton
and MacKenty 1987). Our new data necessarily include broad-band images
for continuum removal, and we use these to further discuss the stellar
populations of the QSO and probable neighbours.

   The observing progam is heavily constrained by practical  
considerations.
First, using the existing narrow-band filters, only certain lines at 
certain redshifts can be observed. Second, as in all programs with the
camera, only QSOs with a bright guide star nearby can be observed.  
Third,
the filter wheel for the observing run cannot hold all available  
filters,
so further downselection is generally needed. Finally, the QSOs  
observable
depended on the scheduled dates of observation.

   The observations were made on two nights in January 2000, and the  
above
constraints left us with 12 targets for emission line imaging. Their
distribution around the sky allowed good observations to be made of 5  
of these,
and broad-band imaging was performed for a further one which had not
previously been observed.  The result is that while we obtained good  
data
on the objects, the sample finally observed had very little latitude  
for
selection by properties such as redshift, luminosity, or radio  
emission.
Nevertheless, a reasonable range of redshifts and radio characteristics  
was
achieved. Table 1 summarizes the observations.

   The objects were observed with the narrow band filter that contains  
the
emission line, and also with the broad-band filter most appropriate for
continuum subtraction. The narrow-band filters used have bandwidths of 
150-200A, and QSO redshifts were selected to place the emission lines  
well
within this - in all cases within the FWHM of the roughly gaussian  
bandpass.
In velocity, our filters correspond to rest wavelength ranges of 2000  
to
4600 km.s$^{-1}$, and in all cases, we observe emission within 650  
km.s$^{-1}$
of (and up to 2000 km.s$^{-1}$ from) the QSO redshift. Table 1 lists  
the 
filter central wavelengths.

   Each set of observations was carried out as a series of
200-300 sec exposures in broad-band and 600 sec exposures in  
narrow-band.
These were generally executed as a set of dithered exposures marking a
square of about 7 arcsec side, sometimes with one in the centre.
Each observation set also included one or more short exposures of the
guide star in the same positions on the detector, to be used as PSF  
data.
Photometric standard stars were observed with the broad-band and one of  
the narrow-band filters as a check on the continuum subtractions.

   The weather was clear on most of both nights, but the image quality  
was
significantly superior on the second night. In some objects, data from  
night
one were found not to add information to the subset of data from night  
2.
Table 1 lists details including image quality and sky brightness during  
the
observing run.

   In the sections below, we discuss the objects and the observations  
on 
each in turn. The data were processed in standard ways, as described in  
paper 
1, using the dithered science exposures to create a sky image. 
Where there was a bright guide star
in the frame, the bright signal pixels were masked out when creating  
the median 
sky frame. The final images were derived by subtracting the mean sky  
image 
derived in this way, for the filter used. These were then averaged with
sigma clipping after being shifted to remove the dither offsets. The  
shifts
were measured from centroids of bright objects in the images.

   There are several ways to estimate the scaling to apply between the
broad and narrow band filter images, for continuum removal. In order of
increasing reliability, these include measuring the signal rates from  
dome 
flat exposures, calculating the relative throughput from the bandpasses  
and
filter transmissions, observations of standard or guide stars, and 
measurements of stars or foreground galaxies in the final science  
images. 
We did not obtain bright star exposures for all the filters, but the  
science
images contained objects that were seen in both filters in all cases.  
In the
sections below, we comment on the process as it applied in the  
different cases.
In all cases, the continuum and narrow-band images were obtained close  
in time,
on the same night, to minimise changes in seeing and image correction
from the AO system. Where observations were obtained on both nights,  
continuum
and narrow-band images were processed before being combined or
compared. Thus, PSF modelling and subtraction are not a critical part  
of the
emission line results reported.

   We did additionally perform some PSF-subtraction, mainly in the  
broad-band 
QSO images. In general, this process involves matching the QSO PSF by  
using a 
short exposure on the guide star, which is then smoothed and scaled so  
that the 
nucleus is still the brightest part of the result. Detailed modelling  
of the 
degradation of the PSF with angle from the guide star (see Hutchings et  
al 1998) 
was not attempted, since it is not adequately calibrated by the  
exposures
in this program, which was aimed at the narrow-band imaging.  

We discuss each case individually below. In individual sections below,  
we also
comment on objects that have intrinsic
blue colours. By this, we mean that rest wavelength colours can arise  
only
from a young population of stars, dominated by spectral types O and B.
The QSO broad-band magnitudes are accurate to 0.1m, from the  
photometric
calibration and expected photon errors. Table 2 summarizes the  
principal 
results from all program objects, described in detail in the following  
sections.

\section{PKS 0348-120}

    This QSO at z = 1.52 is a compact flat spectrum radio source with  
flux
about 500 mJy (Preston et al 1985, Neff et al 1989), and our VLA 
A-configuration 5 GHz map from the latter paper shows no structure. 
The visible magnitude is 19.0 and Wright et al (1983) give J magnitude  
17.7 
and K magnitude 16.1.

   The field was observed in H band and a nearby narrow-band filter  
which
includes the redshifted H$\alpha$. While the field was 
observed on both nights, the  images
from night 1 were the worst of the run, and those from night 2 were 3  
times
sharper in FWHM. Thus, almost all our information comes from the second
night's data. Figures 1 and 2 show the images of this QSO.

   The QSO is clearly resolved in our images (see profile plots in  
Figure 3).
The QSO also  has a slightly curved narrow jet-like extension, seen
clearly in the broad-band image at position angle $\sim$-120$^o$.
The narrow-band image has a sharper linear feature closer in to the  
nucleus,
with similar position angle ($\sim$-135$^o$). While narrow, this feature
has FWHM close to the PSF, and its ridge extends along several pixels. 
In both the raw and PSF-subtracted images, the continuum feature appears
separated from the central object. The H-band feature is  visible 
out to some 0.6" from the QSO, while the narrow-band feature is seen 
within 0.15" of the nucleus. 
 We measure the QSO H band magnitude as 17.1, which is consistent with 
the J and K band values quoted above. In addition to the jet-like  
features, 
the emission line image has
asymmetrical extended flux to the NW and the S while the H band image
is more nearly azimuthally symmetrical. The PSF-subtracted H band image  
shows
extension along the NE-SW direction on both sides of the nucleus, with  
the
`jet' being part of this structure. The jet feature may not connect  
with 
the main host galaxy in the subtracted image, and thus might be
a tidally extended companion being accreted. These results are robust  
to
all feasible PSF scaling factors in the subtraction, as the features  
lie
well outside the central PSF.

    The total flux of the host galaxy thus
derived is 1/7 of the total, which corresponds to a luminosity of
M$_H$ -24.6 for the host galaxy, including the outer extended regions.
The emission line equivalent width ($H\alpha$ plus [N II]) is 300A from  
the resolved narrow-band flux.

The field around the QSO has a few faint 
objects in it, with a range of sizes and fluxes. Aside from the guide  
star, 
there is a star of comparable brightness to the QSO, a nucleated round  
galaxy, 
a faint diffuse galaxy, and 3 resolved objects within about 2 arcsec of  
the QSO.

   What we have found for this QSO is that it is resolved in both  
continuum
and line emission. The H band resolved light extends over one arcsec  
and
is elongated about the nucleus, including the linear feature seen in the 
raw image.
This `host galaxy' flux corresponds to a luminous galaxy at the QSO
redshift. The H$\alpha$ line emission is less extended but also suggests
a narrow feature within the host galaxy, and more diffuse light that  
spreads
orthogonal to the host galaxy.  We speculate that the
host galaxy is linear, or edge-on, possibly merging, and undergoing a  
burst 
of star-formation which is visible as H$\alpha$ emission in material  
blown away from the plane of the galaxy. 
The unresolved nuclear radio source suggests the QSO event is
relatively young: see Neff and Hutchings (1990) for an overview of  
radio quasar evolution with redshift. 

\section{0552+398}

    This QSO was studied in paper 1 in K band, with the
same instrumentation. It was reobserved this time in H band and a  
narrow-band
filter to include redshifted [O~III]. As we noted in our earlier paper,  
the
QSO lies in a rich field of objects that could be companions. The QSO  
is
a compact GHz-peaked radio source (O'Dea et al 1990), which is just
resolved as a VLBI source (Schilizzi and Shaver 1981), with a halo
of major axis of PA 110$^o$ to 70$^o$. At higher frequencies the source  
is
better resolved and may be undergoing superluminal expansion. 
The radio source is powerful and is considered to show synchrotron
self-absorption and radio recombination lines (Bell et al 1984). All  
these
properties suggest the source is immersed in a dense local medium and  
probably
young. 

  In paper 1 we noted the high density of companions likely to be  
associated
with this QSO. Thus, objects in the field may also have [O~III]  
emission in 
the narrow passband, and may not be relied on to define the null ratio  
of 
narrow and broad-band signals. The filter passbands and throughput  
values
indicate that the ratio in our images should be 0.09 or less. Of the 6
objects with measurable narrow band signal, four have values near 0.07,  
so we have adopted this as the null ratio. The QSO (plus host) has a  
value 
of 0.092, which indicates an equivalent width of [O~III] of 35A. The  
other
object measured in the NB field has a value of 0.11 (EW=40A), and may  
indeed
be an emission-line galaxy at the QSO redshift. This object
lies some 15" to the NW of the QSO. Uncertainties
in the above measures amount to 1$\sigma$ errors in EW of some 5A,  
based
purely on the difference in the broad to narrow band image fluxes.

   The continuum-subtracted [O~III] image shows an elongated structure
on both sides of the nucleus with position angle very close to that of  
the
VLBI radio source(s) (see Figures 1 and 4). Note that the closely spaced contours
near the nucleus arise from some PSF mismatch near the nucleus. The 
contours along the NW-SE extensions are, however, robust to a wide range
of PSF scalings and locations. The emission is stronger to the  
NW, and 
is seen over a distance of about 0.5". Using a short exposure of the  
guide star 
as PSF, the H-band image with PSF subtracted shows some faint light in 
a broad region to the NW (i.e. also along the axis of
radio and line emission), that is invariant to the details of the PSF
subtraction. This extends to some 0.7" from the nucleus. The overall
host galaxy light is strongly dependent on the PSF scaling and  
modelling
and lies in the range 3 to 10 times fainter than the QSO. The QSO is  
very
luminous, with M$_H$ at -30.1 (using our cosmology parameters from  
paper 1 
of H$_0$=75 km.s$^{-1}$.mpc$^{-1}$, q$_0$=0.5, and no  
K-correction). This 
means the host galaxy absolute magnitude is in the range -28 to -29,  
which 
is also extremely luminous.

   The QSO colour can be derived by using the measures from paper 1.  
While 
there are minor uncertainties because of weather differences, the  
differences
between K' and K calibrations, and the possibility of
variability of the QSO, we find that the QSO is significantly reddened.
The QSO is in fact the reddest object in the field. This may indicate
internal dust associated with extensive star-formation, and may
be the reason that the [O III] equivalent width is the lowest we detect
(see table 2), if the dust is patchy and associated with the emission  
line regions.  (Note that the red colour of the QSO and location of
the NB passband will cause us to overestimate the
EW, although only at the few percent level.) 
This scenario is consistent with the youth and 
local absorption noted above from the radio observations.

 The new H band images, combined with the data from paper 1, reveal the
colours of many of the nearby objects, which are of interest.   
 Our earlier K band images and the new H band ones do not overlap  
exactly, 
so it is impossible to get
colours for all objects in paper 1. We do see the same objects in
the overlapping areas of old and new data. The colours we measure range  
over
some 0.8 mag in H-K, and magnitudes range from 2 to 6 magnitudes  
fainter than
the QSO. All objects are compact and although the brightest may be  
foreground
objects, the rest are unusually luminous and blue objects if they are  
at the 
QSO redshift.

   Overall, this QSO may be in a central luminous galaxy of a  
group of
compact bright galaxies. There is line emission aligned with the compact  
radio source that is more extended than the radio, but lies within the
body of the host galaxy. It seems probable that this QSO is in its  
early stages
and lies within a reddened galaxy with considerable star-formation. It  
should
be detectable as a source further into the infrared.

\section{MG 0828+2919}

   This QSO appears in radio source catalogues as 7C0825+2930,  
B0825+295,
and J0828+29. The radio flux is given as 719 mJy at 151 MHz (7C)
(Willott et al 1998), 226 mJy at 1.4 GHz (White and Becker 1992),
100 mJy at 8 GHz (Wilkinson et al 1998), and 50-100 mJy 
at 5 GHz (Falco et al 1998). The spectral index is given by Willott et  
al
as 0.45, and Wilkinson et al report a faint jet 1" long at position
angle -90$^o$ to -60$^o$. The optical redshift determination is  
reported
at 2.315 by Willott et al and 2.322 by Falco et al. Either redshift
puts emission lines well within the filters used.

The optical identification is not described in detail, and there are two
faint compact objects within about 2 arcsec. The radio source position
lies within 0.2 arcsec of the southern of the two objects.
   Our observations show that both these objects are very compact, and
their azimuthal profiles are similar in the H and K' band images. 
 
Figure 3 shows the H-band profiles, in which the Southern object is designated
QSO, and it is slightly more extended than the PSF and the other  
object.
In two dimensions, the extensions in the southern object are 
the same in H and K band images from both nights. Subtraction of our  
PSF 
profile scaled to remove the central peak leaves faint extended flux,
largely to the NE, in the southern object. Figures 1, 4 and 5 show images 
of this QSO and its companion.

   The narrow band flux is larger in the southern object, in both  
[O~III]
and H$\alpha$, which confirms our identification. If the [O~III]  
continuum
subtraction from the narrow-band image is scaled to remove the northern  
object, the QSO image has a nuclear
source plus a possible extension to the NE, about 0.3 arcsec long. This
subtraction is not perfect, however, since it either leaves a point  
source 
at the centre of the northern
object, or oversubtracts its extended light. This probably arises from
changes in the seeing between the continuum and narrow-band images: if  
so,
it also provides a calibration of the effect, assuming there is no  
[O~III]
emission in the northern object. The scale factor that we derive this  
way 
is exactly what we expect based on the bandpasses of the two filters  
and
exposure times used: 7.4. However, this is not a robust result as the
transparency during the observations was variable. 

   The narrow band image that isolates H$\alpha$ shows a larger  
difference
between the two objects, but unfortunately the northern object lies on  
the
intersection of two quadrants of the detector, so that it is not
a good control for the scaling of the continuum
image subtraction. However, using the value derived from the filter
bandpasses and exposure times, the signal left in the northern object  
is 
within the noise level of zero, while leaving a substantial signal at  
the
southern object. The morphology of the continuum-subtracted 
QSO image is extended in all directions, particularly along NE-SW. 
We note that both [O III] and H$\alpha$ are extended to the W.
The stronger NE extension seen in [O~III] is not apparent in H$\alpha$.

    The excess light in the narrow-band images compared with the scaled
broad-band images corresponds to equivalent widths overall of 120A in
[O~III] and 290A in H$\alpha$. The H$\alpha$ filter will also include
light from the [N II] nebular lines in the host galaxy. The H and K'
magnitudes for the QSO and northern object are 17.8, 17.3 and 17.8,  
17.8
respectively, based on our standard star data, again with the caveat  
that 
the transparency was variable. Willott et al (1998) report a K  
magnitude of
17.55 and Falco et al (1998) report I magnitude of 18.5.

    The only other object in the 35" field, apart from the guide star,
is a galaxy about 7.9" to the south of the QSO. It is smooth and round  
(like 
an elliptical) of diameter some 1.5", and has H and K' magnitudes 18.0  
and 17.5.
The close grouping and similar magnitudes of the objects detected  
raises 
the question of whether they are related. However, the galaxy is too  
bright to
be at the QSO redshift, and is certainly a foreground object, probably  
at
redshift 0.5 or lower from its magnitude, colour, and size. The  
unresolved
nature, colour, and lack of line emission from the northern companion
all suggest that it is a star, of type mid-F or earlier.

    Thus, the information on the QSO from our data is that its host  
galaxy
is marginally detected as light extending to the NE. Its luminosity  
indicates
M$_H$ -23 or so (with uncertainty of a magnitude). There is line  
emission from
both [O~III] and H$\alpha$. The [O~III] may be extended along a `jet'  
whose
direction is close to that of the reported radio jet, but much shorter.  
The
H$\alpha$ emission appears to be generally extended along the NE-SW  
directions,
which suggests it is located throughout the host galaxy. All these  
results
are close to the detection limit of our data, so need confirmation with
much larger signal levels, and high spatial resolution.

\section{0915-213}

This QSO is a compact radio source,
and is the lowest redshift object in this sample at z = 0.85. In
paper 1 we noted a knot 1.4" to the NW and a faint wisp of luminosity
extending beyond, particularly in K band, in the direction of an object
(considered to be a star, 6" from the QSO). We note that the published 
visible magnitude for the QSO of 17.5 in Hewitt and Burbidge (1993)
appears to be discrepant, as their references give 18.5. 

Our new data are in J band, in which our narrow-band filter isolates
redshifted H$\alpha$. The images had the lowest FWHM of any we  
obtained,
close to 0.11", in both narrow and broad band filters. The ratio of narrow
to broad band filter signal is 0.081 for the 3 brightest objects
in the field, and 0.23 for the QSO. With 0.08 scaling, the continuum
subtracted images removed the nearby star completely but left a
considerable signal at the QSO. However, the continuum-subtracted QSO
image also has FWHM 0.11", so the H$\alpha$ emission is also close to
unresolved, and must arise overwhelmingly in or very near the nucleus. 
The subtracted image is elongated slightly NE-SW (see Figure 1), so 
there may be some marginally extended line emission. The equivalent  
width 
of the H$\alpha$ (plus [N II]) implied is 290A. This yields a continuum  
J
magnitude of 17.3 (17.1 without the H$\alpha$ correction).

  The knot 1.3" to the NW, which is clearly seen in the H and K band  
images,
is much fainter in J band, but present. Its J magnitude is poorly  
determined
at about 22. The dither pattern used in the observation and the  
position of 
the guide star in this field resulted in some unevenness in the sky  
images
in the region between the QSO and the nearby star. 
It was not possible to detect the extended flux seen in the H and
K bands from paper 1 (taken with different dither pattern), and its  
J-band 
flux would be very low if it is as red as the knot. There is no 
evidence for H$\alpha$ emission from this region in our data.

  Accurate PSF-subtraction from the J band image requires detailed  
modelling not  
possible with the data from this run. An approximate subtraction was  
done 
of a bright star closer to the guide star (profiles shown in Figure 3).  
This results in an elongation NE-SW, similar
to that of the continuum-subtracted H$\alpha$ image. While this is in  
the
general direction of the guide star (the direction the system might
produce distortions), it is different by some 30$^o$. The flux after 
PSF-subtraction corresponds to J magnitude 20.2 for the host galaxy,  
which 
is similar to the values found in H and K in our earlier paper,  
although less
accurate with estimated error at 0.3 mag. 

   Overall, the new evidence on this object is that the host galaxy is  
blue
(NIR colour indices close to zero) and some  times fainter than the  
nucleus, 
which corresponds to an absolute
magnitude close to -22 in J, H, and K with no K-correction. It is  
possibly
extended NE-SW in both continuum and H$\alpha$. The H$\alpha$ is  
predominantly
unresolved, and there is no extended line emission associated with the  
wisp
seen in H and K bands, to the NW. The nearby companion or knot is  
redder
but detected in J band, and not in H$\alpha$.

\section{1227+4641}

   This is a QSO from the Hamburg quasar survey. Only its redshift  
(2.154)
and visible magnitude (18.4)  are published (Engels et al 1998), and it  
is 
not a known radio source. Its redshifted [O~III] is in our narrow-band  
filter. 
We find no source in the FIRST radio survey at this position,
to a flux density of $\sim$1 mJy.

   The H band magnitude of the QSO is 15.8. The 2.6 mag difference to  
V = 18.4
corresponds to rest wavelengths of 1700 to 5000. This suggests that the  
QSO 
has UV extinction, contains a significant red component, or has varied.  
The guide star provides a rough check on the continuum subtraction  
scale 
factor at about 0.066, while the filter bandpass data suggest a value  
of 0.071. 
The observed ratio of 0.133
for the QSO corresponds to an [O~III] equivalent width of about 150A.
The redshifted C IV 1550A EW indicated by the published spectrum of 
Engels et al (1998) is about 100, and the flux out to observed 6500A 
is declining with increasing wavelength.

   The continuum-subtracted [O~III] image shows a broad asymmetry of  
emission
to the SW (see Figures 1 and 6), with a central jet-like feature that  
curves to
the S from the nucleus, some 0.3" long. Such structures are generally 
associated with radio jet activity from the nucleus. However, because of the  
very small
scale and spatial resolution limits, it is possible that the structure  
arises
from the merging or
superposition of two bright emission line regions. The continuum image  
has a
small inner bright extension from the nucleus in the same direction,  
but no
signature of galaxy interaction.

   The removal of the broad-band PSF is not straightforward. There is
a companion star to the guide star that serves well as PSF, and it has
considerably sharper profile. The QSO image is extended along the line
to the guide star, and so needs modelling. If the PSF image is  
elongated
so as to match the QSO radial profile, it loses the diffraction rings,
which can be seen in the QSO image. Figures 3 and 6 illustrate this
object. In any subtraction, there is significant
residual light for any operation that leaves a light maximum at the QSO
nuclear position (i.e. is not oversubtracted). There is no definite or
persistent structure for the PSF-removed image, and the signal level  
that
remains indicates nuclear to host galaxy flux ratio in the range 2.5
to close to 1.0. This is consistent with the well-resolved [O~III]  
flux,
and indicates a host galaxy of M$_H$-28.4, which is very luminous, even  
with
the large uncertainty of 0.5 magnitudes in this value.

 The H-band image shows little in the field other than the QSO. The
position of the guide star just outside the field of view, leads to  
several 
diffuse reflections near the QSO, which are easily recognized by their 
round flat-topped profiles. In addition, a bright diffraction spike for  
the
guide star runs through the QSO image. It is unlikely that these  
spurious
features hide any real objects, and we detect only one compact galaxy,  
close 
to the guide star. 

   Overall, we find evidence for a luminous host galaxy, with  
asymmetrical
line emission. The object should be re-observed with different camera
orientation and in different filters, to learn more about it, and to  
confirm the
results. 

\section{12487+5706}

   This QSO was observed only in H band, on both nights, as its  
redshift
does not place an emission line in a NB filter bandpass. The image
quality was comparable and the data were combined. There is no source  
at
the QSO position in the FIRST radio survey, to a limit of about 2mJy.  
The
QSO was identified from the Einstein EMSS X-ray survey as a 19th  
magnitude
QSO at redshift 1.843.

 Our broad-band data show the
QSO to have H magnitude 19.0, which suggests it may be very blue, if
it has not varied. The
field is empty except for one faint galaxy some 17" to the south, and a  
large
foreground spiral galaxy 12" to the north.   

   The QSO appears resolved in good seeing (see Figure 3). 
The image shows the elongation
along the line to the guide star, as also seen in by 1227+4641. The  
host galaxy
flux for our best subtraction of profile and image (i.e. not  
oversubtracted
in the inner regions) is 1.8 mag fainter
than the QSO, which corresponds to an absolute magnitude of -23.9.
The elongation towards the guide star is asymmetrical and thus may be  
partly
instrinsic to the QSO host galaxy, but there is no definite structure
in the raw or PSF-subtracted image. The uncertainty is about 0.5  
magnitudes 
for the host galaxy magnitude, but in any case it
is less luminous than the others in this program. This appears to be a
radio-quiet QSO in a poor environment.

\section{Discussion}

   This investigation has detected extended line emission in all of the
five QSOs observed, and the results are summarized in Table 2.
The line emission is very compact and requires $\sim$0.1"
resolution to be seen. Three of the QSOs have redshifts close to 2.3,
and the others are lower, at 0.9 and 1.5. All but one of the QSOs are
(very compact) radio sources, and in the two radio sources that are  
resolved,
the line emission is aligned with it. We do not detect line emission  
from
outside the QSO host galaxy, such as found by Stockton and MacKenty  
(1987)
in a low redshift sample, which include similar emission line fluxes.
However, they note, along with Boroson, Persson, and Oke (1985) that  
strong 
extended [O~III] emission is generally found in QSOs with extended  
steep-spectrum
radio emission, and our sample are all compact radio sources or radio  
quiet.
As well as being at high redshift, the QSOs in this program are  
generally of 
higher luminosity.

  We have not attempted to compare the emission line region sizes and  
fluxes
in Table 2, because of the effects of redshift and varying image  
quality and
signal level, as well as small sample size. 
The H$\alpha$ equivalent width is similar for the 3 objects
where it is detected, and arises in all cases in extended regions
that probably indicate young stars, and certainly arises within the  
confines 
of a normal-sized host galaxy. H$\alpha$ also appears as a jet-like  
nuclear
feature in 0348-120, that may also indicate nuclear beaming. The [O  
III]
emission in all three detected cases forms a radial linear feature from  
the
nucleus that suggests beamed nuclear activity, particularly as it  
aligns
with the radio structure in the two resolved radio sources. The [O III]
equivalent width has a wider range than H$\alpha$, and we have already
speculated that the cause of the one low value (0552+398) may be dust.
The apparent sizes of the emission line regions (see Figure 1) are  
quite
similar, with an inverse correlation with redshift at the half-power  
level
(hence similar in actual size). All these quantites need
to be investigated in a larger sample before we can expect significant
correlations.

   The success of this program indicates that it is possible to study
the evolution of emission line gas in QSO host galaxies, as well as  
their stellar 
populations and morphology. The linear nature of the emission line  
regions,
and the association with known radio structure, strongly suggest that  
line
emission is excited by nuclear radiation, which is beamed. The sizes of  
the
regions we have observed lie between the NELR structure seen in low
redshift Seyferts and in low redshift QSOs, but we have not sampled the  
full range of radio structures, which are strongly correlated with ELR
at low redshifts. Since the high redshift QSOs are in host galaxies at
early stages of their formation and evolution, these observations will
provide a measure of the beaming, gas density, and effects of merging
within the host galaxies at different cosmic times. Thus, extending the  
redshift range, and inclusion in the sample of large radio 
source QSOs are highly desirable in order to map the evolution of the  
ELR
of QSOs.  

   Two of our objects are very bright in H and (hence) very red:  
0552+398
and 1227+4641. Both appear to have considerable resolved H-band flux,
which needs more careful PSF modelling, since the implied luminosities
are very high. It is possible that these are young and dusty objects
in which we speculate that much of the nuclear flux may be scattered  
into 
the NIR.

    Table 2 includes
counts of other objects in the field: these are not sensitive to the
range of exposure times or image quality in the sample, as they can all
be seen within the range of image quality in the data, and in exposure
subsets as short as our shortest observations.  The numbers given 
exclude the guide star and other clearly identified stars, and large 
foreground galaxies. On the other hand, faint objects can be lost near  
to 
the guide star, and it is not always easy to distinguish faint stars  
from 
compact galaxies.  The fields for the two objects also in paper 1 are  
not 
placed exactly as before, so there are small differences in galaxy  
counts 
for that reason too. We have already
discussed this in the overlapping sample in paper 1, in some more  
detail,
but the numbers are consistent and the range of richness of possible
companions in the field is large enough to be worthy of comment. 

   The QSO 0552+398 still stands out as having an unusually rich  
environment,
and the two radio-quiet QSOs have poor environments (but so does one  
radio-loud
one: 0828+2919). There are suggestions that 3 of the QSOs have visible  
signs
of galaxy merging: 0348-120, 0915-213, and 1227+4641, but deeper  
broad-band
imaging is required. The broad-band exposures in this program were  
designed
principally to enable continuum-subtraction from the narrow-band  
images.
It is clear that a proper study of high redshift host galaxies and  
companions
requires very deep exposures as well as high resolution.

   The narrow-band imaging results need to be extended to a sample of
QSOs of the type that have extended line emission at low redshift:  
steep
spectrum extended radio sources, those with strong narrow emission  
lines, 
and weak Fe II emission. 

     We thank Eric Steinbring for some of the image processing software
and discussions on AOB data. We also thank the referee for a careful
and useful reading of the paper.

\footnotesize
\begin{deluxetable}{lllrcclrrcc} 
\tablenum{1}
\tablecaption{Journal of observations}
\tablehead{\colhead{QSO} &\colhead{z} &\colhead{m$_v$} &\colhead{m$_v$}
&\colhead{offset} &\colhead{Filter} &\colhead{Line} &\colhead{exp}
&\colhead{sky\tablenotemark{a}} &\colhead{FWHM} &\colhead{Night}\\
&&&\colhead{GS} &\colhead{GS} &\colhead{$\lambda$(A)} &&\colhead{(sec)}  
&\colhead{(300s)}
&\colhead{(")}  }
\startdata
0348-120 &1.52 &19.0 &12.3 &20" &H &Cont &800 &1800 &0.50 &1\nl
&&&&&&&2400 &3000 &0.14 &2\nl
&&&&&16590 &H$\alpha$ &2400 &140 &0.40 &1\nl
&&&&&&&3000 &200 &0.13 &2\nl
0552+398 &2.37 &18.0 &11.7 &27" &H &Cont &1500 &1400 &0.30 &1\nl
&&&&&&&1200 &2200 &0.11 &2\nl
&&&&&16900 &[O~III] &4800 &150 &0.25 &1\nl
&&&&&&&3000 &140 &0.12 &2\nl
0828+2919 &2.32 &18.5 &13.3 &13" &H &Cont &1200 &1050 &0.17 &1\nl
&&&&&&&1200 &1900 &0.21 &2\nl
&&&&&K' &Cont &1200 &2300 &0.18 &1\nl
&&&&&16590 &[O~III] &4800 &140 &0.16 &1\nl
&&&&&&&2400 &145 &0.14 &2\nl
&&&&&21660 &H$\alpha$ &2400 &145 &0.18 &1\nl
&&&&&&&2400 &190 &0.15 &2\nl
0915-213 &0.85 &18.5 &11.9 &12" &J &Cont &2400 &360 &0.11 &2\nl
&&&&&12070 &H$\alpha$ &5400 &70 &0.11 &2\nl
1227+4641 &2.15 &18.4 &11.0 &15" &H &Cont &1200 &500 &0.35 &1\nl
&&&&&&&1200 &1300 &0.15 &2\nl
&&&&&15750 &[O~III] &4800 &80 &0.35 &1\nl
&&&&&&&3000 &70 &0.15 &2\nl
12487+5706 &1.84 &19.0 &11.2 &20" &H &Cont &1500 &800 &0.19 &1\nl
&&&&&&&1600 &800 &0.17 &2\nl
\enddata
\tablenotetext{a}{sky count level in 300 secs exposure}
\end{deluxetable}
\normalsize

\footnotesize
\begin{deluxetable}{lrrcrrccl} 
\tablenum{2}
\tablecaption{QSO properties}
\tablehead{\colhead{QSO} &\colhead{5GHz} &\colhead{z} &\colhead{m$_v$}
&\colhead{m$_H$} &\colhead{line EW\tablenotemark{a}}
&\colhead{\#\tablenotemark{b}}
&\colhead{N/H\tablenotemark{c}} &\colhead{Comment (M$_H$ host)}\\
&\colhead{(mJy)} &&&&\colhead{(\AA)}  }
\startdata
0348-120 &400 &1.52 &19.0 &17.1 &H$\alpha$ ~300 &3 &7 &J=17.7,  
K=16.1, Symm
host\nl
&&&&&&&&H$\alpha$ `jet', asymm, (-25)\nl
\nl
0552+398 &4900 &2.37 &18.0 &15.3 &[O~III] 35 &11 &(6) &Red, host  
resolved,
(-28)\nl 
&&&&&&&&VLBI GHz peak, [O~III] aligned\nl
\nl
0828+2919 &100 &2.32 &18.5 &17.9 &H$\alpha$ ~290 &1 &20 &Host resolved  
(-23),
H$\alpha$ broad\nl
&&&&K' 17.8 &[O~III] 120 &&&[O~III] along radio jet\nl
\nl
0915-213 &600 &0.85 &18.5 &16.7 &H$\alpha$ ~290 &4 &15&Host resolved  
(-22) 
+ knot\nl
&&&(17.5?) &J 17.3 &&&&H$\alpha$ nuc + E-W structure\nl
\nl
1227+4641 &$<$1 &2.15 &18.4 &15.8 &[O~III] 150 &1 &3 &Red; symm host  
resolved
(-28)\nl
&&&&&&&&[O~III] curved jet + asymm region\nl
\nl
12487+5706 &$<$2 &1.84 &19.0 &19.0 &$-$ &1 &(5) &Probably resolved  
(-24)\nl
\enddata
\tablenotetext{a}{in observed wavelength frame}
\tablenotetext{b}{number of possible galaxy companions in 35" field}
\tablenotetext{c}{ratio of unresolved to resolved H band flux}
\end{deluxetable}

\normalsize

\newpage
\centerline{References}
\vskip 20pt
Bell M.B., Seaquist E.R., Mebold U., Reif K., and Shaver P., 1984,  
A\&A, 130, 1

Boroson T.A., Persson S.E., and Oke J.B., 1985, ApJ, 293, 1 

Engels D., Hagen H-J., Cordis L., Kohler S., Wisotzki L., Reimers D.,  
1998,
A\&ASup, 128, 507 

Falco E.E., Kockanek C.S., Munoz J.A., 1998, ApJ, 494, 47

Hewitt A., and Burbidge G., 1993, ApJS, 87, 1

Hutchings J.B., Crampton D., Morris S.L., Steinbring E., 1998, PASP,  
110, 374

Hutchings J.B., Crampton D., Morris S.L., Durand D., 1999, AJ, 117,  
1109 
(paper 1)

Neff S.G., Hutchings J.B., and Gower A.C., 1989, AJ, 97, 1291

Neff S.G., and Hutchings J.B., 1990, AJ, 100, 1441

O'Dea C.P., Baum S.A., Stanghellini C., Morris G.B., Patniak A.R.,
Gopal-Krishna, 1990, A\&AS, 84, 549

Preston R.A., et al, 1985, AJ, 90, 1599

Schilizzi R.T., and Shaver P., 1981, A\&A, 96, 365

Stockton A., and MacKenty J.W., 1987, ApJ, 316, 584

White R.L. and Becker R.H., 1992, ApJS, 79, 331

Willott C.J., Rawlings S., Blundell K.M., Lacy M., 1998, MNRAS, 300,  
625

Wilkinson P.N., Browne I.W.A., Patniak A.R., Wrobel J.M., *** B.S.,  
1998,
MNRAS, 300, 790

Wright A.E., Ables J.G., Allen D.A., 1983, MNRAS, 5, 793

\newpage

\centerline{Captions to figures}

1. Contour plots of continuum-subtracted narrow-band images. QSO nuclei
are situated at the coordinate origin. North is up and east to the  
left.
Successive contour intervals increase roughly by a factor 1.5. 

2. Grey scale images of 0348-120. N is up and E to the left, and images  
are
2.1" on a side. A: H band, B: H$\alpha$ narrow-band, C:  
continuum-subtracted
H$\alpha$, D: PSF-subtracted H band. Note continuum 
extended flux to SW and H$\alpha$ SW `jet' and extended flux to NW.

3. Azimuthally averaged profile plots of QSOs and PSFs. Filters as  
labelled.

4. Grey scale continuum-subtracted narrow-band images, 1.6" on a side.
Left: 0552+398 [O III]. Right: 0828+2919 [O III].

5. Grey scale images of 0828+2919. N is up and E to the left, and  
images
are 4.1" on a side. QSO is southern object. A: summed H and K' band  
images,
B: H$\alpha$ continuum-subtracted image, C: PSF-subtracted H-band (QSO  
only), 
D: [O III] continuum-subtracted image.

6. Grey scale images of 1227+4641. N is up and E to the left, and  
images
are 2.8" on a side. A: H band, B: [O~III] narrow-band, C:  
continuum-subtracted
[O~III], D: PSF-subtracted H band. 

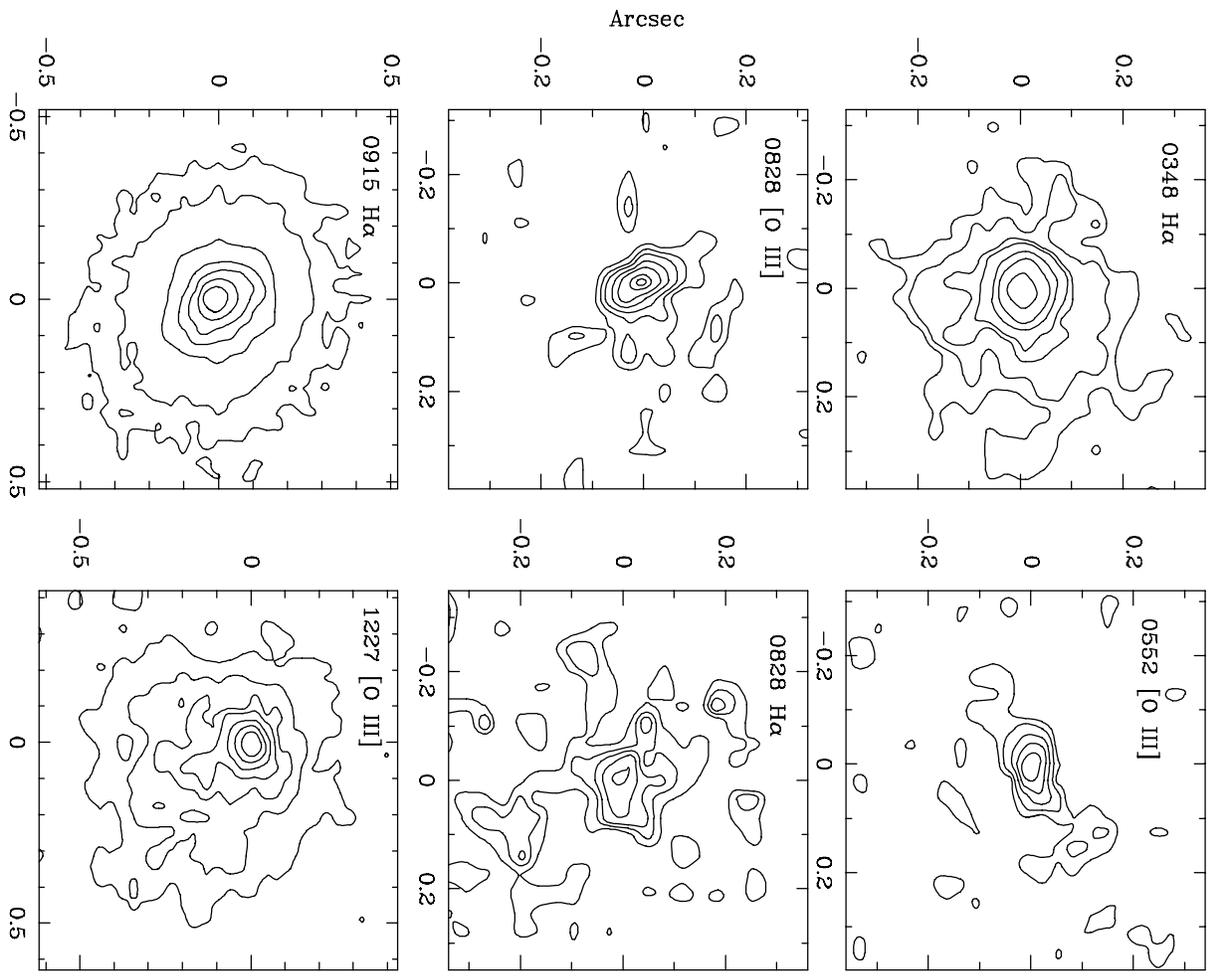
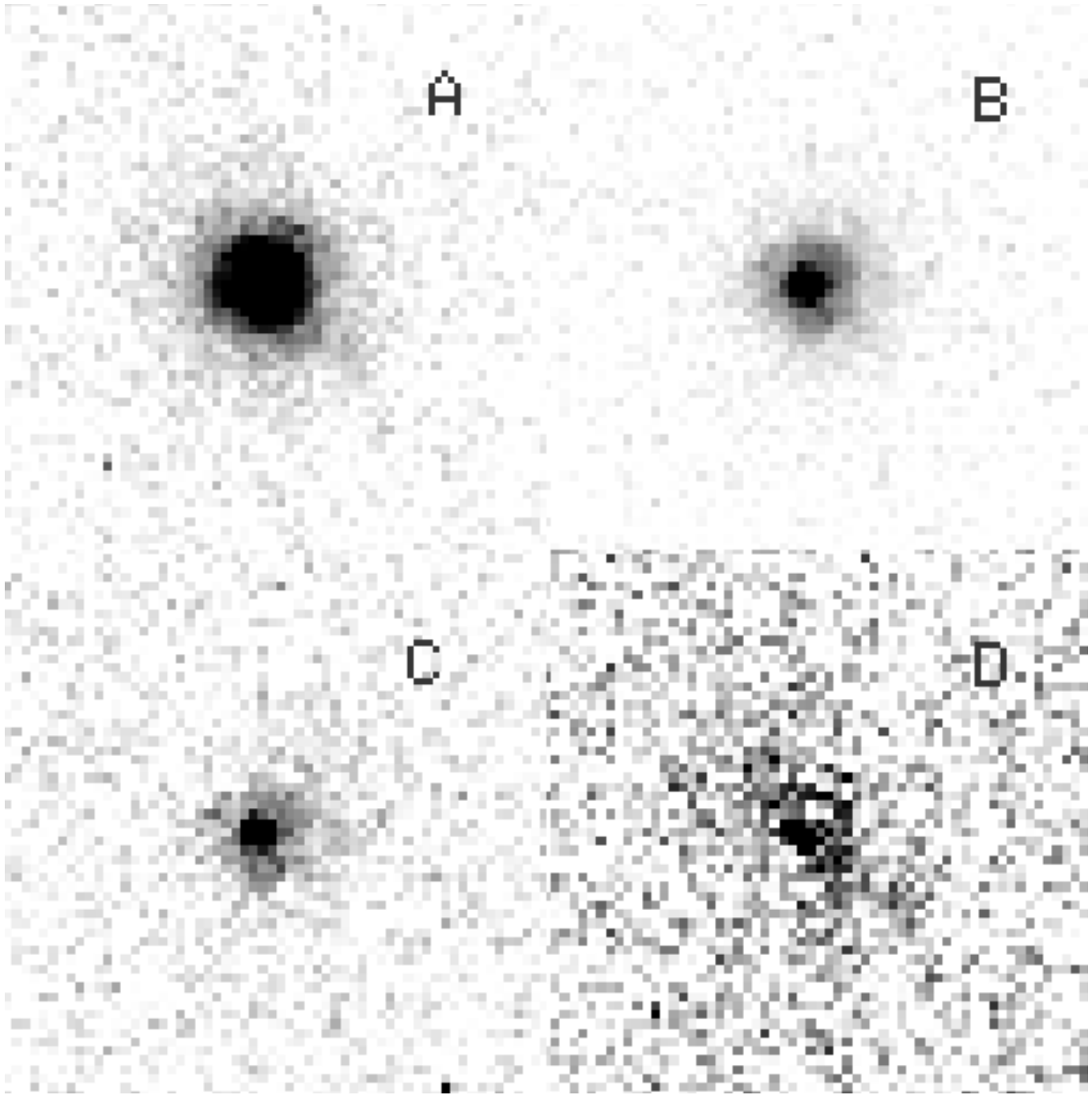
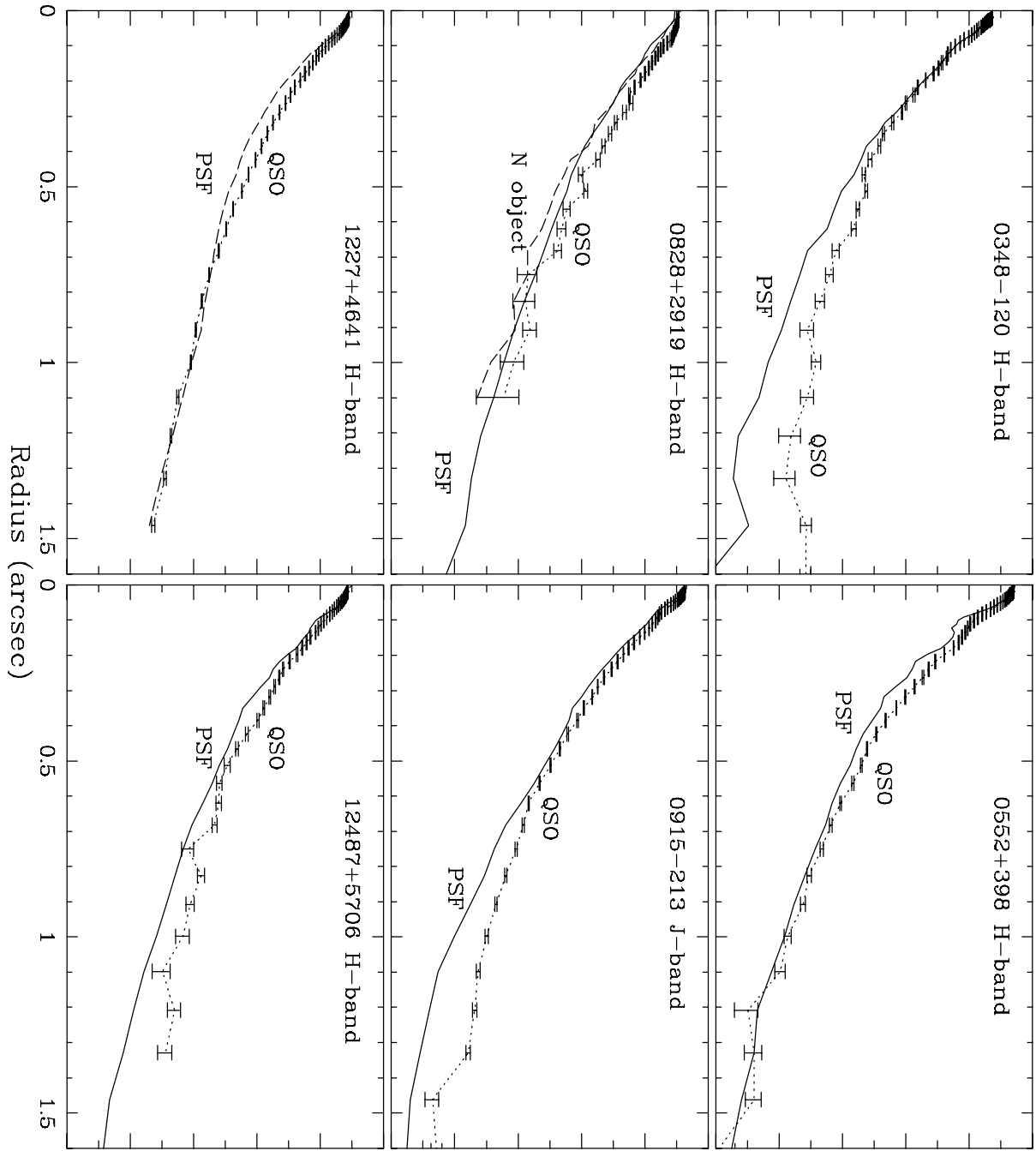
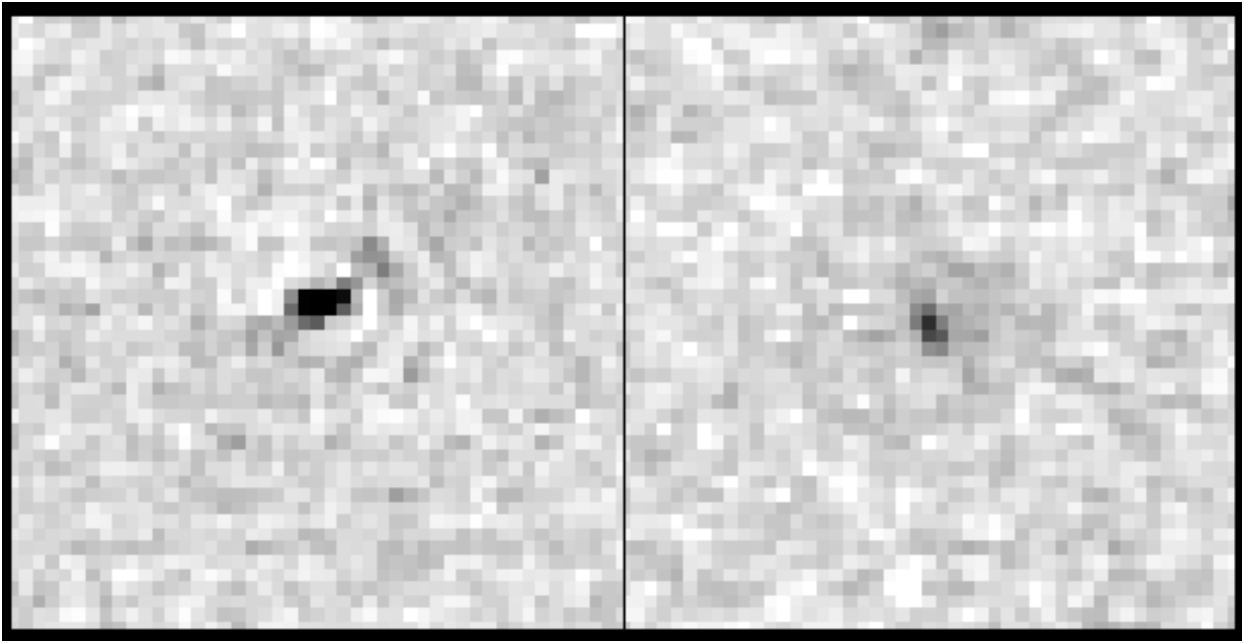
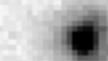
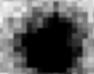
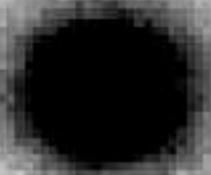
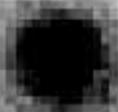
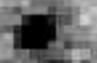
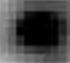
\end{document}